\documentstyle[aps]{revtex}

\begin{document}
\title{Final State Interactions and $\Delta S=-1,$ $\Delta C=\pm 1$ $B$--decays}
\author{Fayyazuddin}
\address{National Centre for Physics and Physics Department, Quaid-i-Azam University,%
\\
Islamabad, Pakistan.}
\maketitle

\begin{abstract}
The final state interactions (FSI) in $\Delta S=-1$, $\Delta C=\pm 1$ decays
of $B$-meson are discussed. The rescattering corrections are found to be of
order of $15-20\%$. The strong interaction phase shifts are estimated and
their effects on $CP$- asymmetry are discussed.
\end{abstract}

\section{Introduction}

Direct $CP$-violation in the decays $B\rightarrow f$ and $\bar{B}\rightarrow 
\bar{f}$ depends on the strong final state interactions. In fact the $CP$%
-asymmetry parameter vanishes in the limit of no strong phase shifts. The
purpose of this paper is to study the $\Delta C=\pm 1,$ $\Delta S=-1$ $B$%
-decays taking into account the final state interactions \cite
{f1,f2,f3,f4,f5}. Such decays are described by the effective Lagrangians 
\begin{eqnarray}
{\cal L}_{eff} &=&\frac{G_F}{\sqrt{2}}V_{cb}V_{us}^{*}[\bar{s}\gamma ^\mu
(1-\gamma _5)u][\bar{c}\gamma _\mu (1-\gamma _5)b]  \nonumber \\
{\cal L}_{eff} &=&\frac{G_F}{\sqrt{2}}V_{ub}V_{cs}^{*}[\bar{s}\gamma ^\mu
(1-\gamma _5)c][\bar{u}\gamma _\mu (1-\gamma _5)b]  \label{01}
\end{eqnarray}
Both these Lagrangians have $\Delta I=\frac 12$. The weak phase in the
Wolfenstein parameterization \cite{f6} of CKM matrix \cite{f7} is given by 
\begin{equation}
\frac{V_{ub}V_{cs}^{*}}{V_{cb}V_{us}^{*}}=\sqrt{\rho ^2+\eta ^2}e^{i\gamma
},\qquad \sqrt{\rho ^2+\eta ^2}=0.36\pm 0.09  \label{02}
\end{equation}
It is quite difficult to reliably estimate the final state interactions in
weak decays. The problem is somewhat simplified by using isospin and $SU(3)$
symmetry in discussing the strong interaction effects. We will fully make
use of these symmetries. Moreover we note in Regge phenomenology, the strong
interactions scattering amplitudes can be written in terms of Pomeron
exchange and exchange of $\rho -A_2$ and $\omega -f$ trajectories in $t$%
-channel. The problem is further simplified if there is an exchange
degeneracy. In fact this is the case here. In $s(u)$ channel only the states
with quark structure $s\bar{c}$ $(c\bar{s})$ can be exchanged ($s,$ $u,$ $t,$
are Mandelstam variables). An important consequence of this is that since $%
\bar{K}D$ has quantum numbers $C=1,$ $S=-1$, only a state with structure $c%
\bar{s}$ ($C=1,$ $S=1,$ $Q=+1)$ can be exchanged in $u$-channel where as no
exchange is allowed in the $s$-channel (exotic). On the other hand $\bar{K}%
\bar{D}$ ($C=-1,$ $S=-1$) state is non-exotic and we can have an exchange of
a state with structure $s\bar{c}$ ($C=-1,$ $S=-1,$ $Q=-1$) in $s$-channel
and a state with structure $d\bar{c}$ or $u\bar{c}$ ($C=-1,$ $S=0$) in $u$%
-channel for a quasi-elastic channel such as $\bar{K}^0\bar{D}^0\rightarrow
\pi ^{+}D_s^{-}$ or $K^{-}\bar{D}^0$($\bar{K}^0D^{-}$)$\rightarrow \pi
^0D_s^{-}$. In Regge Phenomenology, exotic $u$-channel implies exchange
degeneracy i.e. in $t$-channel $\rho -A_2$ and $\omega -f$ trajectories are
exchange degenerate. Taking into account $\rho -\omega $ degeneracy, all the
elastic or quasielastic scattering amplitudes can be expressed in terms of
two amplitudes which we denote by $F_P$, $F_\rho $ and $\bar{F}_\rho
=e^{-i\pi \alpha \left( t\right) }F_\rho $, where $F_P$ is given by pomeron
exchange, $F_\rho $ by particle exchange trajectory for which $\alpha _\rho
\left( t\right) =\alpha _{A_2}\left( t\right) =\alpha _\omega \left(
t\right) =\alpha _f\left( t\right) =\alpha \left( t\right) $. We have argued
above, that elastic and quasi-elastic scattering amplitudes can be
calculated fairly accurately. This combined with the following physical
picture \cite{f8,f9} gives us a fairly reliable method to estimate the
effect of rescattering on weak decays. In the weak decays of $B$-measons,
the $b$ quark is converted into $b\rightarrow c+q+\bar{q}$, $b\rightarrow
u+q+\bar{q}$; since for the \cite{f8,f9} tree graph the configuration is
such that $q$ and $\bar{q}$ essentially go together into the color singlet
state with the third quark recoiling, there is a significant probability
that the system will hadronize as a two body final state. This physical
picture has been put on a strong theoretical basis in \cite{f10,r1}. In this
picture the strong phase shifts are expected to be small at least for tree
amplitude.

Now the discontinuity or imaginary part of decay amplitude is given by \cite
{r2,f11} 
\begin{equation}
\mathop{\rm Im}%
A_f=\sum_nM_{fn}^{*}A_n  \label{03}
\end{equation}
where $M_{fn}$ is the scattering amplitude for $f\rightarrow n$. According
to above picture the important contribution to the decay amplitude $%
A_f\equiv A\left( B\rightarrow f\right) $ in Eq. (\ref{03}) is from those
two body decays of $B$ which proceed through tree graphs. Thus it is
reasonable to assume that the decays proceeding through the following chains 
\begin{eqnarray}
\bar{B}^0 &\rightarrow &K^{-}D^{+}\rightarrow \bar{K}^0D^0  \nonumber \\
\bar{B}_s^0 &\rightarrow &K^{-}D_s^{+}\rightarrow \pi ^{-}D^{+}  \label{04}
\end{eqnarray}
and 
\begin{eqnarray}
\bar{B}^0 &\rightarrow &\pi ^{+}D_s^{-}\rightarrow \bar{K}^0\bar{D}^0 
\nonumber \\
B^{-} &\rightarrow &\pi ^0D_s^{-}\rightarrow \bar{K}^0D^{-}  \nonumber \\
&\rightarrow &\eta D_s^{-}\rightarrow K^{-}\bar{D}^0  \nonumber \\
\bar{B}_s^0 &\rightarrow &K^{+}D_s^{-}\rightarrow \pi ^{+}D^{-}  \label{05}
\end{eqnarray}
may make a significant contribution to the decay amplitudes. Note that the
intermediate channels are those channels for which the decay amplitude is
given by tree amplitude for which the factorization anstaz is on a strong
footing \cite{r1,r2}.

In view of above arguments, the dominant contribution in Eq. (\ref{03}) is
from a state $n=\acute{f}$ where the decay amplitude $A_{\acute{f}}$ is
given by the tree graph. For the quasi elastic channels listed in Eqs. (\ref
{04}) and (\ref{05}), the scattering amplitude $M_{f\acute{f}}$ is given by
the $\rho $-exchange amplitude $F_\rho $ or $\bar{F}_\rho $. The purpose of
this paper is to calculate the rescattering correction to $A_f$ by the
procedure outlined above. Final state interactions considered from this
point of view will be labelled as FSI (A). In an alternative point of view,
labelled (B) one lumps all channels other the elastic channel in one
category. In the random phase approximation of reference \cite{f10}, one may
take the parameter $\rho =\left[ \overline{\left| A_n^2\right| }\right]
^{1/2}/\left. A_f\right. ,$ $n\neq f$ nearly $1$ for the color suppressed
decays but $\rho $ small for the two body decays dominated by tree graphs.
In this paper the observational effects of final state interactions will be
analyzed from the point of view (A).

\section{Decay Amplitudes Decomposition}

The amplitudes for various $\bar{B}$ decays are listed in Table1. They are
characterized according to decay topologies: (1) a color-favored tree
amplitude $T$, (2) a color-suppressed tree amplitude $C$, (3) an exchange
amplitude $E$, and (4) an annihilation amplitude $A$. The isospin
decomposition of relevant decays along with strong interaction phases are
also given in Table1. Isospins and $SU(3)$ symmetry gives the following
relationships for various decay amplitudes 
\begin{eqnarray}
A_{-+}+A_{00} &=&A_{-0}  \nonumber \\
B_{-s^{+}}-B_{-+} &=&A_{-+}  \label{61} \\
\bar{A}_{-0}-\bar{A}_{0-} &=&\bar{A}_{00}  \nonumber \\
\sqrt{2}\bar{A}_{0s}-\sqrt{6}\bar{A}_{8s^{-}} &=&2\bar{A}_{0-}  \nonumber \\
\bar{B}_{+s^{-}}-\bar{B}_{+-} &=&\bar{A}_{+s^{-}}  \nonumber \\
\bar{B}_{+-}-\sqrt{6}\bar{B}_{80} &=&2\bar{A}_{00}  \label{07}
\end{eqnarray}
First we note that in the naive factorization ansatz, we have the following
relations between the amplitudes of various topologies\cite{f12} 
\begin{eqnarray}
\frac CT &=&\left( \frac{a_2}{a_1}\right) \frac{%
f_DF_0^{B-K}(m_D^2)(m_B^2-m_K^2)}{f_KF_0^{B-D}(m_K^2)(m_B^2-m_D^2)} 
\nonumber \\
&\approx &\left( \frac{a_2}{a_1}\right) (0\cdot 72)  \label{08} \\
\frac ET &=&\left( \frac{a_2}{a_1}\right) \frac{%
f_{B_s}F_0^{D_s-K}(m_B^2)(m_{D_s}^2-m_K^2)}{%
f_KF_0^{B_s-D_s}(m_K^2)(m_{B_s}^2-m_{D_s}^2)}  \nonumber \\
&\approx &\left( \frac{a_2}{a_1}\right) (0\cdot 08)  \label{09} \\
\frac{\bar{T}}T &\approx &0\cdot 72\sqrt{\rho ^2+\eta ^2}  \label{10}
\end{eqnarray}
\begin{mathletters}
\label{101}
\begin{eqnarray}
\frac{\bar{C}}T &=&\left( \frac{a_2}{a_1}\right) \frac{%
f_DF_0^{B-K}(m_D^2)(m_B^2-m_K^2)}{f_{D_s}F_0^{B-\pi }(m_{D_s}^2)(m_B^2-m_\pi
^2)}  \nonumber \\
&\approx &\left( \frac{a_2}{a_1}\right)  \label{11a} \\
\frac{\bar{A}}{\bar{T}} &=&\frac{f_BF_0^{D-K}(m_B^2)(m_D^2-m_K^2)}{%
f_{D_s}F_0^{B-\pi }(m_{D_s}^2)(m_B^2-m_\pi ^2)}  \nonumber \\
&\approx &0\cdot 08  \label{11b}
\end{eqnarray}
where $a_2/a_1\approx 0.2-0.3$. The numerical values have been obtained
using $f_D\approx 200$ MeV, $f_{D_s}\approx 240$ MeV, $f_K\approx 158$ MeV, $%
f_B\approx 180$ MeV. 
\end{mathletters}
\begin{eqnarray}
\frac{F_0^{B-K}\left( m_D^2\right) }{F_0^{B-D}\left( m_K^2\right) } &\approx
&0.05\approx \frac{F_0^{D_s-K}\left( m_B^2\right) }{F_0^{B_s-D_s}\left(
m_K^2\right) }  \nonumber \\
F_0^{B-K}\left( m_D^2\right) &\approx &F_0^{B-\pi }\left( m_{D_s}^2\right)
\label{12}
\end{eqnarray}
Since the amplitudes $C$ ($\bar{C}$), $E$, $\bar{A}$ are suppressed relative
to tree amplitude, they are subject to important corrections due to
rescattering.

\section{Rescattering}

In order to calculate rescattering corrections and to obtain $s$-wave strong
phases, we consider the scattering processes 
\begin{eqnarray*}
P_a+\bar{D} &\rightarrow &P_b+\bar{D} \\
P_a+D &\rightarrow &P_b+D
\end{eqnarray*}
where $P_a$ is a pseudoscalar octet. Using $SU(3)$, the scattering amplitude
can be written as 
\begin{equation}
M=\chi ^{\dagger }\left[ \bar{F}_1\frac 12[\lambda _b,\lambda _a]+\bar{F}_2%
\frac 12\{\lambda _b,\lambda _a\}+\bar{F}_3\delta _{ba}\right] \chi
\label{13}
\end{equation}
where $\chi $ is an $SU(3)$ triplet 
\begin{equation}
\chi =\left( 
\begin{array}{l}
\bar{D}^0 \\ 
D^{-} \\ 
D_s^{-}
\end{array}
\right)  \label{14}
\end{equation}
For the process $P_a+D\rightarrow P_b+D$, we replace $\bar{F}_i$ ($i=1,2,3$)
by $F_i$ and $\chi $ by the triplet 
\begin{equation}
\chi =\left( 
\begin{array}{l}
D^0 \\ 
D^{+} \\ 
D_s^{+}
\end{array}
\right)  \label{15}
\end{equation}
In order to express the scattering amplitudes in terms of Regge
trajectories, it is convenient to define two amplitudes 
\begin{eqnarray}
M^{+} &=&P+f+A_2  \nonumber \\
&=&-C_P\frac{e^{-i\pi \alpha _P(t)/2}}{\sin \pi \alpha _P(t)/2}\left( \frac s%
{s_0}\right) ^{\alpha _P(t)}+\left[ -C_f\frac{1+e^{-i\pi \alpha _f(t)}}{\sin
\pi \alpha _f(t)}\left( \frac s{s_0}\right) ^{\alpha _f(t)}-C_{A_2}\frac{%
1+e^{-i\pi \alpha _{A_2}(t)}}{\sin \pi \alpha _{A_2}(t)}\left( \frac s{s_0}%
\right) ^{\alpha _{A_2}(t)}\right]  \label{16} \\
M^{-} &=&\rho +\omega =\left[ C_\omega \frac{1-e^{-i\pi \alpha _\omega (t)}}{%
\sin \pi \alpha _\omega (t)}\left( \frac s{s_0}\right) ^{\alpha _\omega
(t)}+C_\rho \frac{1+e^{-i\pi \alpha _\rho (t)}}{\sin \pi \alpha _\rho (t)}%
\left( \frac s{s_0}\right) ^{\alpha _\rho (t)}\right]  \label{17}
\end{eqnarray}
Due to exchange degeneracy, for linear Regge trajectories 
\begin{eqnarray}
\alpha _\rho \left( t\right) &=&\alpha _{A_2}\left( t\right) =\alpha _\omega
\left( t\right) =\alpha _f\left( t\right) =\alpha _0\left( t\right) +\acute{%
\alpha}t  \label{18} \\
C_f &=&C_\omega ;\qquad C_{A_2}=C_\rho  \nonumber \\
C_\omega &=&C_\rho  \label{19}
\end{eqnarray}
We take\cite{f11,f13} $\alpha _0=0.44\approx 1/2$ and $\acute{\alpha}=0.94$%
GeV$^{-2}\approx 1$GeV and for the pomeron $\alpha _P\left( t\right) =\alpha
_P\left( 0\right) +\acute{\alpha}_Pt$, $\alpha _P\left( 0\right)
=1.08\approx 1$ and $\acute{\alpha}_P\approx 0.25$GeV$^{-2}$. In particular
for the processes $K^{-}D^0\rightarrow K^{-}D^0$ and $K^{-}\bar{D}%
^0\rightarrow K^{-}\bar{D}^0$,we get 
\begin{eqnarray}
M(K^{-}D^0 &\rightarrow &K^{-}D^0)=P+(f-\omega )+(A_2-\rho )  \nonumber \\
&=&iC_P\left( \frac s{s_0}\right) e^{bt}-2(C_\omega +C_\rho )\frac 1{\sin
\pi \alpha (t)}\left( \frac s{s_0}\right) ^{\alpha (t)}  \nonumber \\
&=&iC_P\left( \frac s{s_0}\right) e^{bt}-4C_\rho \frac 1{\sin \pi \alpha (t)}%
\left( \frac s{s_0}\right) ^{\alpha (t)}  \nonumber \\
&=&F_P+F_\rho  \label{20}
\end{eqnarray}
\newline
\begin{eqnarray}
M(K^{-}\bar{D}^0 &\rightarrow &K^{-}\bar{D}^0)=P+(f+\omega )+(A_2+\rho ) 
\nonumber \\
&=&F_P+e^{-i\pi \alpha (t)}F_\rho =F_P+\bar{F}_\rho  \label{21}
\end{eqnarray}
where $b=\acute{\alpha}_P\ln (s/s_0)$. We take \cite{r2} $C_P=5$. In order
to estimate $C_\rho $, we note that $SU(3)$ gives 
\begin{eqnarray}
\gamma _{\rho K^{+}K^{-}} &=&-\gamma _{\rho K^0\bar{K}^0}=\frac 12\gamma
_{\rho \pi ^{+}\pi ^{-}}=\frac 12\gamma _0  \nonumber \\
\gamma _{\omega K^{+}K^{-}} &=&\gamma _{\omega K^0\bar{K}^0}=\frac 12\gamma
_0  \nonumber \\
\gamma _{\rho D^{+}D^{-}} &=&-\gamma _{\rho D^0\bar{D}^0}=-\gamma _{\omega
D^{+}D^{-}}=-\gamma _{\omega D^0\bar{D}^0}  \label{22}
\end{eqnarray}
For our purpose, we will take $\gamma _{\rho D^{+}D^{-}}\approx \gamma
_{\rho K^{+}K^{-}}=\frac 12\gamma _0$, so that 
\begin{equation}
C_\rho =\gamma _{\rho K^{+}K^{-}}\gamma _{\rho D^0\bar{D}^0}=-\frac 14\gamma
_0^2  \label{23}
\end{equation}
For $\gamma _0^2$, we use that value $\gamma _0^2=72$ as given in reference 
\cite{f13}.

Now using Eq. (\ref{13}) and Eqs. (\ref{20}) and (\ref{21}), we can express
all the elastic or quasi-elastic scattering amplitudes in terms of the Regge
amplitudes $F_P$ and $F_\rho $. These amplitudes are given in TableII.

We are now in a position to discuss the rescattering corrections to the
decay amplitudes. From Eq. (\ref{03}), the two particle unitarity gives \cite
{f11,f13}.

\begin{eqnarray}
DiscA(\bar{B}^0 &\rightarrow &\bar{K}^0D^0)=\frac 1{32\pi }\frac{\left| \vec{%
p}\right| }s\int d\Omega M^{*}(\bar{K}^0D^0\rightarrow K^{-}D^{+})A(\bar{B}%
^0\rightarrow K^{-}D^{+})  \nonumber \\
&\approx &\frac 1{16\pi s}\int_{-2\left| \vec{p}\right| ^2}^0dtM^{*}(\bar{K}%
^0D^0\rightarrow K^{-}D^{+})A(\bar{B}^0\rightarrow K^{-}D^{+})
\end{eqnarray}
where we have put $\left| \vec{p}\right| \simeq \frac 12\sqrt{s}$. Now using
TableII and Eqs. (\ref{20}) and (\ref{23}), we get 
\begin{eqnarray}
DiscA(\bar{B}^0 &\rightarrow &\bar{K}^0D^0)=\gamma _0^2\frac 1{16\pi s}\frac{%
A(\bar{B}^0\rightarrow K^{-}D^{+})}{\sin \pi \alpha _0}\times \left(
s/s_0\right) ^{\alpha _0}\int_{-2\left| \vec{p}\right| ^2}^0dte^{t\acute{%
\alpha}\ln s/s_0}  \nonumber \\
&=&\frac{\gamma _0^2}{16\pi }\left( s/s_0\right) ^{\alpha _0-1}\frac 1{\ln
(s/s_0)}A(\bar{B}^0\rightarrow K^{-}D^{+})  \label{24}
\end{eqnarray}
where in evaluating the integral in Eq. (\ref{24}), we have put $\sin \pi
\alpha (t)=\sin \pi \alpha _0=\sin \frac \pi 2=1$ and $\acute{\alpha}s_0=1$
i.e. $s_0=1$GeV$^2.$

We now use dispersion relation \cite{f11,f13,f14} to obtain 
\begin{equation}
A(\bar{B}^0\rightarrow \bar{K}^0D^0)_{FSI}=\frac{\gamma _0^2}{16\pi }\frac{A(%
\bar{B}^0\rightarrow K^{-}D^{+})}{\ln (m_B^2/s_0)}\frac{\sqrt{s_0}}{m_B}%
\frac 1\pi \int_{\left( m_D+m_K\right) ^2}^\infty \left( s/s_0\right)
^{\alpha _0-1}\frac{ds}{s-m_B^2}  \label{25}
\end{equation}
where in $\ln (s/s_0)$, we have put $s=m_B^2$. Noting that $\alpha _0\approx
1/2$, we get 
\begin{eqnarray}
A(\bar{B}^0 &\rightarrow &\bar{K}^0D^0)_{FSI}=\frac{\gamma _0^2}{16\pi }%
\frac 1{\ln (m_B^2/s_0)}\frac{\sqrt{s_0}}{m_B}\frac 1\pi \left[ i\pi +\ln 
\frac{1+x}{1-x}\right] A(\bar{B}^0\rightarrow K^{-}D^{+})  \nonumber \\
&\equiv &\epsilon e^{i\theta }A(\bar{B}^0\rightarrow K^{-}D^{+})  \label{26}
\end{eqnarray}
where 
\begin{eqnarray}
x &=&\frac{m_D+m_K}{m_B}\simeq 0\cdot 447,m_B=5\cdot 279  \nonumber \\
\epsilon &=&\frac{\gamma _0^2}{16\pi }\frac 1{\ln (m_B^2/s_0)}\frac{\sqrt{s_0%
}}{m_B}\left[ 1+\frac 1{\pi ^2}\left( \ln \frac{1+x}{1-x}\right) ^2\right]
^{1/2}  \nonumber \\
&=&\gamma _0^2\left[ 1\cdot 18\times 10^{-3}\right] \simeq 0\cdot 08 
\nonumber \\
\theta &=&\tan ^{-1}\left[ \frac \pi {\ln \frac{1+x}{1-x}}\right] \simeq 73^0
\label{27}
\end{eqnarray}
\mbox{$>$}%
>From Eqs. (\ref{61}, \ref{07}) and TableII, following the same procedure, we
can eaisly calculate the rescattering corrections for other decays. Hence
after taking into account rescattering corrections to the decay amplitudes,
we get 
\begin{mathletters}
\label{28}
\begin{eqnarray}
A_{00} &=&a_{00}e^{i\delta _{00}}+\epsilon e^{i\theta }a_{-+}e^{i\delta
_{-+}}  \label{28a} \\
A_{-+} &=&a_{-+}e^{i\delta _{-+}}  \label{28b} \\
A_{-0} &=&a_{-0}e^{i\delta _{-0}}+\epsilon e^{i\theta }a_{-+}e^{i\delta
_{-+}}  \label{28c} \\
B_{-s^{+}} &=&b_se^{i\delta _s}  \label{28d} \\
B_{-+} &=&b_{1/2}e^{i\delta _{1/2}}+\frac 12(1+i)\epsilon e^{i\theta
}b_se^{i\delta _s}  \label{28e}
\end{eqnarray}
\end{mathletters}
\begin{mathletters}
\label{29}
\begin{eqnarray}
\bar{A}_{00} &=&(\bar{a}_{00}e^{i\bar{\delta}_{00}}+\epsilon e^{i\theta }%
\bar{a}e^{i\bar{\delta}})e^{i\gamma }  \label{29a} \\
\bar{A}_{-0} &=&(\bar{a}_{-0}e^{i\bar{\delta}_{-0}}+\frac 12\epsilon (1-%
\frac i3)\bar{a}e^{i\bar{\delta}})e^{i\gamma }  \label{29b} \\
\bar{A}_{0-} &=&(\bar{a}_{0-}e^{i\bar{\delta}_{0-}}-\frac 12\epsilon (1+%
\frac i3)\bar{a}e^{i\bar{\delta}})e^{i\gamma }  \label{29c} \\
B_{+s^{-}} &=&\bar{b}_se^{i\bar{\delta}_s}  \label{29d} \\
\bar{B}_{-+} &=&\bar{b}_{1/2}e^{i\bar{\delta}_{1/2}}+\frac 12(1+i)\epsilon
e^{i\theta }\bar{b}_se^{i\bar{\delta}_s}  \label{29e}
\end{eqnarray}
The phase factor $i$ arises due to the phase factor $e^{i\pi \alpha (t)}$ ($%
\bar{F}_\rho =\ e^{i\pi \alpha (t)}F)$ $[$we can also write $(1+i)\frac 12=%
\frac 1{\sqrt{2}}e^{i\frac \pi 4},$ $(1\pm \frac i3)=\frac{\sqrt{10}}3e^{\pm
i\phi },$ $\phi =\tan ^{-1}1/3=18^0]$

\section{Strong interaction phase shifts}

For the $s$-wave scattering, the $l=0$ partial wave scattering amplitude $f$
is given by 
\end{mathletters}
\begin{equation}
f=\frac 1{16\pi s}\int_{-4\vec{p}^2}^0M(s,t)dt  \label{30}
\end{equation}
Hence from Eqs. (\ref{20}) and (\ref{21}), we get for $\sqrt{s}=5\cdot 279$%
GeV, 
\begin{eqnarray}
f_P &=&\frac 1{16\pi s}\frac{iC_P}b\left( \frac s{s_0}\right) \simeq 0.12i
\label{31} \\
f_\rho &=&\frac{\gamma _0^2}{16\pi }\frac 1{\ln \left( \frac s{s_0}\right) }%
\left( \frac s{s_0}\right) ^{1/2}\simeq 0.08  \label{32} \\
\bar{f}_\rho &=&\frac{\gamma _0^2}{16\pi }(-i)\frac 1{\ln \left( \frac s{s_0}%
\right) -i\pi }\left( \frac s{s_0}\right) ^{1/2}\simeq 0.04+0.04i  \label{33}
\end{eqnarray}
Using Table II and Eqs. (\ref{31}-\ref{33}), we can determine the $s$-wave
scattering amplitude $f$ and $S$-matrix, $S=1+2if\equiv \eta e^{2i\Delta }$
for each individual scattering process. They are given in Table III.

Now, using Eq. (\ref{05}) we get 
\begin{equation}
\mathop{\rm Re}%
A_f\left( 1-\eta e^{-2i\Delta }\right) -i%
\mathop{\rm Im}%
A_f\left( 1+\eta e^{2i\Delta }\right) =\frac 1i\sum_{n\neq f}A_nS_{nf}^{*}
\label{34}
\end{equation}
Taking the absolute square of Eq. (\ref{34}), we obtain, writing $A_f=\left|
A_f\right| e^{i\delta _f}$: 
\begin{equation}
\left| A_f\right| ^2\left[ \left( 1+\eta ^2\right) -2\eta \cos 2\left(
\delta _f-\theta \right) \right] =\sum_{n,\acute{n}\neq f}A_nS_{nf}^{*}A_{%
\acute{n}}^{*}S_{\acute{n}f}  \label{35}
\end{equation}
Using the random phase approximation of reference\cite{r2}: 
\begin{equation}
\sum_{n,\acute{n}\neq f}A_nS_{nf}^{*}A_{\acute{n}}^{*}S_{\acute{n}%
f}=\sum_{n\neq f}\left| A_n\right| ^2\left| S_{nf}\right| ^2=\overline{%
\left| A_n\right| }^2\left( 1-\eta ^2\right)  \label{36}
\end{equation}
we obtain from Eqs. (\ref{35}) and (\ref{36}): 
\begin{mathletters}
\label{37}
\begin{eqnarray}
\tan ^2\left( \delta _f-\Delta \right) &=&\left( \frac{1-\eta }{1+\eta }%
\right) \frac{\left[ \rho ^2-\left( \frac{1-\eta }{1+\eta }\right) \right] }{%
\left[ 1-\rho ^2\left( \frac{1-\eta }{1+\eta }\right) \right] }  \label{37a}
\\
\rho ^2 &=&\frac{\overline{\left| A_n\right| }^2}{\left| A_f\right| ^2},%
\frac{1-\eta }{1+\eta }\leq \rho ^2\leq \frac{1+\eta }{1-\eta }  \label{37b}
\end{eqnarray}
Except for the parameter $\rho $, everything is known from the TableIII. For
the moment let us evaluate the final state phase shifts $\delta _f$ from Eq.
(\ref{37}) for three values of $\rho ^2$ viz $\rho ^2=\frac{\left( 1-\eta
\right) }{\left( 1+\eta \right) },$ $0\cdot 25$ and$1$. These phase shifts
are tabulated in Table IV for various decays.

The following remarks are in order. Irrespective of parameter $\rho $, it
follows from the above analysis that 
\end{mathletters}
\begin{eqnarray}
\bar{\delta}_s &=&\delta _s  \nonumber \\
\bar{\delta}_{1/2} &=&\delta _{1/2}  \label{39}
\end{eqnarray}
It is reasonable to assume that $\rho $ is of the same order for all the
favored decays; it may be different for the suppressed decays but within a
set it does not differ much. Under this assumption it follows from Table IV 
\begin{eqnarray}
\bar{\delta} &=&\delta _{-+}  \nonumber \\
\bar{\delta}_{-0} &=&\bar{\delta}_{0-}  \nonumber \\
&&\bar{\delta}_{00=}\delta _{00}  \label{40}
\end{eqnarray}
Further if we take the point of view consistent with that of reference \cite
{f8} i.e. for the decay amplitudes dominated by tree graphs the final state
interactions are negligible, ($\rho $: small) then we may conclude 
\begin{eqnarray}
\delta _{-+} &=&\bar{\delta}\leq 7^0  \nonumber \\
\delta _{-0} &\leq &13^0  \nonumber \\
\delta _s &=&\bar{\delta}_s\leq 10^0  \label{41}
\end{eqnarray}
To proceed further, we first consider the case (A): For the color suppressed
decays, since we have already taken into account the rescattering
corrections c.f. Eqs. (\ref{28}) and (\ref{29}), it is reasonable to take
the same value of $\rho $ for all the decays ($\rho \ll 1$). Then from Table
IV, we have 
\begin{eqnarray}
\bar{\delta}_{00=}\delta _{00} &=&\delta _{-+}\leq 7^0  \nonumber \\
\bar{\delta}_{-0} &=&\bar{\delta}_{0-}\leq 10^0  \nonumber \\
\bar{\delta}_{1/2} &=&\delta _{1/2}\leq 10^0  \label{42}
\end{eqnarray}
For the case (B): The phase shifts for color favored decays will be small as
given in Eq. (\ref{41}); but for the color suppressed decays $\rho $ would
be of the order $1$ and the phase shifts for these decays will be in the
range of $20^0$ as given in Table IV. For this case, in Eqs. (\ref{28}) and (%
\ref{29}), $\epsilon $ will be taken to zero as the effect of rescattering
is supposed to be reflected in the parameter $\rho =1$.

\section{Observational Consequences of final state interactions (FSI)}

We note that the decay amplitudes after taking into account FSI are given in
Eqs. (\ref{28}) and (\ref{29}). From these equations, we obtain (neglecting
terms of order $\epsilon ^2$) 
\begin{mathletters}
\label{43}
\begin{eqnarray}
a_{-0}^2 &=&a_{-+}^2+a_{00}^2+2a_{-+}a_{00}\cos \left( \delta _{-+}-\delta
_{00}\right)  \label{43a} \\
\Gamma \left( B^{-}\rightarrow K^{-}D^0\right) &=&a_{-0}^2+2\epsilon
a_{-0}a_{-+}\cos \left( \theta +\delta _{-+}-\delta _{-0}\right)  \label{43b}
\\
\Gamma \left( \bar{B}^0\rightarrow \bar{K}^0D^0\right) &=&a_{00}^2+2\epsilon
a_{00}a_{-+}\cos \left( \theta +\delta _{-+}-\delta _{00}\right)  \label{43c}
\\
\Gamma \left( \bar{B}^0\rightarrow K^{-}D^{+}\right) &=&a_{-+}^2  \label{43d}
\end{eqnarray}

\end{mathletters}
\begin{mathletters}
\label{44}
\begin{eqnarray}
\bar{a}_{-0}^2 &=&\bar{a}_{00}^2+\bar{a}_{0-}^2+2\bar{a}_{00}\bar{a}%
_{0-}\cos \left( \bar{\delta}_{00}-\bar{\delta}_{0-}\right)  \label{44a} \\
\Gamma \left( B^{-}\rightarrow K^{-}\bar{D}^0\right) &\simeq &\bar{a}_{-0}^2+%
\frac{\sqrt{10}}3\epsilon \bar{a}_{-0}\bar{a}\cos \left( \theta -\phi +\bar{%
\delta}-\bar{\delta}_{-0}\right)  \label{44b} \\
\Gamma \left( B^{-}\rightarrow \bar{K}^0D^{-}\right) &\simeq &\bar{a}_{0-}^2-%
\frac{\sqrt{10}}3\epsilon \bar{a}_{0-}\bar{a}\cos \left( \theta +\phi +\bar{%
\delta}-\bar{\delta}_{0-}\right)  \label{44c} \\
\Gamma \left( \bar{B}^0\rightarrow \bar{K}^0\bar{D}^0\right) &\simeq &\bar{a}%
_{00}^2+2\epsilon \bar{a}_{00}\bar{a}\cos \left( \theta +\bar{\delta}-\bar{%
\delta}_{00}\right)  \label{44d} \\
\Gamma \left( B^{-}\rightarrow \pi ^0D_s^{-}\right) &=&\bar{a}^2  \label{44e}
\end{eqnarray}
First we note from Eqs. (\ref{28}), (\ref{29}), (\ref{43a}) (\ref{44a}) that
in the absence of rescattering the amplitudes $\left| A_{-0}\right| $, $%
\left| A_{-+}\right| $, $\left| A_{00}\right| $, and $\left| \bar{A}%
_{-0}\right| $, $\left| \bar{A}_{00}\right| $, $\left| \bar{A}_{0-}\right| $
will form two closed triangles. Any deviation from triangular relations
would indicate rescattering.

Now using the factorization(see TableI) and Eqs. (\ref{08}--\ref{101}), we
obtain $a_{00}/a_{-+}=C/T\approx 0.72\,(a_2/a_1)$, $\bar{a}_{00}/\bar{a}=%
\bar{C}/\bar{T}\approx (a_2/a_1)$, $\bar{a}_{0-}/\bar{a}=\bar{A}/\bar{T}%
\approx 0.08$ and noting that $\epsilon =0.08$, $a_2/a_1\sim \lambda =0\cdot
22$, $\theta =73^0$ and $\phi =18^0$, we note that FSI corrections are in
the range of $15-20\%$, except for $B^{-}\rightarrow \bar{K}^0D^{-}$, (c.f.
Eq. (\ref{43c}) where it is almost zero, since $\theta +\phi \simeq 90^0$.
However we note that the effect of FSI\ is of considerable importance for
the decays $\bar{B}_s^0\rightarrow \pi ^{-}D^{+}$ and $\bar{B}%
_s^0\rightarrow \pi ^{+}D^{-}$ as both these decays in the absence of FSI
are extremely suppressed as these decays occur through $W$-exchange diagram
(c.f. TableI). Taking into account FSI, we obtain from Eqs. (\ref{28}), (\ref
{29}) and (\ref{09}) 
\end{mathletters}
\begin{eqnarray}
\frac{\Gamma \left( \bar{B}_s^0\rightarrow \pi ^{-}D^{+}\right) }{\Gamma
\left( \bar{B}_s^0\rightarrow K^{-}D_s^{+}\right) } &=&\frac{b_{1/2}^2}{b_s^2%
}\left[ 1+\sqrt{2}\epsilon \frac{b_s}{b_{1/2}}\cos \left( \theta +\frac \pi 4%
+\delta _s-\delta _{1/2}\right) +\frac 12\epsilon ^2\frac{b_s^2}{b_{1/2}^2}%
\right]  \nonumber \\
&\simeq &\sqrt{2}\epsilon \left( E/T\right) \cos \left( \theta +\frac \pi 4%
+\delta _s-\delta _{1/2}\right) +\frac 12\epsilon ^2  \nonumber \\
&\simeq &2.3\times 10^{-3}=\frac{\Gamma \left( \bar{B}_s^0\rightarrow \pi
^{+}D^{-}\right) }{\Gamma \left( \bar{B}_s^0\rightarrow K^{+}D_s^{-}\right) }
\label{45}
\end{eqnarray}
In the absence of FSI, this ratio has the value $3\cdot 1\times 10^{-4}$.

We now discuss the effect of FSI on CP-asymmetry. We define 
\begin{eqnarray}
{\cal A}_{\mp } &=&\frac{\Gamma \left( B^{-}\rightarrow K^{-}D_{\mp
}^0\right) -\Gamma \left( B^{+}\rightarrow K^{+}D_{\mp }^0\right) }{\Gamma
\left( \bar{B}^0\rightarrow K^{-}D^{+}\right) }  \nonumber \\
&=&\pm 2\sin \gamma \left[ 
\mathop{\rm Re}%
A_{-0}%
\mathop{\rm Im}%
\bar{A}_{-0}-%
\mathop{\rm Im}%
A_{-0}%
\mathop{\rm Re}%
\bar{A}_{-0}\right] /\left| A_{-+}\right| ^2  \label{46} \\
{\cal R}_{\mp } &=&\frac{\Gamma \left( B^{-}\rightarrow K^{-}D_{\mp
}^0\right) +\Gamma \left( B^{+}\rightarrow K^{+}D_{\mp }^0\right) }{\Gamma
\left( \bar{B}^0\rightarrow K^{-}D^{+}\right) }  \nonumber \\
&=&\left\{ \left| A_{-0}\right| ^2+\left| \bar{A}_{-0}\right| ^2\mp 2\cos
\gamma \left[ 
\mathop{\rm Re}%
A_{-0}%
\mathop{\rm Re}%
\bar{A}_{-0}+%
\mathop{\rm Im}%
A_{-0}%
\mathop{\rm Im}%
\bar{A}_{-0}\right] \right\} /\left| A_{-+}\right| ^2  \label{47}
\end{eqnarray}
where $D_{\mp }^0=\frac{\left( D^0\mp \bar{D}^0\right) }{\sqrt{2}}$ and weak
phase $e^{i\gamma }$ has been taken out. For $\bar{B}^0$ and $B^0$ decays, $%
{\cal A}_{\mp }^0$, ${\cal R}_{\mp }^0$ can be obtained by changing $%
A_{-0}\rightarrow A_{00}$ and $\bar{A}_{-0}\rightarrow \bar{A}_{00}$ in Eqs.
(\ref{46}, \ref{47}). Then using Eqs. (\ref{28}) and (\ref{29}), we get 
\begin{mathletters}
\label{48}
\begin{eqnarray}
{\cal A}_{\mp } &=&\pm 2\sin \gamma \left[ -f\bar{r}\sin (\delta _{-0}-\bar{%
\delta}_{-0})-\epsilon \bar{r}\sin (\theta +\delta _{-+}-\bar{\delta}_{-0})+%
\frac{\sqrt{10}}6\epsilon f\bar{f}\sin (\theta -\phi +\bar{\delta}-\delta
_{-0})\right]  \label{48a} \\
{\cal R}_{-}-{\cal R}_{+} &=&\mp 4\cos \gamma \left[ f\bar{r}\cos (\delta
_{-0}-\bar{\delta}_{-0})+\epsilon \bar{r}\cos (\theta +\delta _{-+}-\bar{%
\delta}_{-0})+\frac{\sqrt{10}}6\epsilon f\bar{f}\cos (\theta -\phi +\bar{%
\delta}-\delta _{-0})\right]  \label{48b}
\end{eqnarray}

\end{mathletters}
\begin{mathletters}
\label{49}
\begin{eqnarray}
{\cal A}_{\mp }^0 &=&\pm 2\sin \gamma \left[ -r_0\bar{r}_0\sin (\delta _{00}-%
\bar{\delta}_{00})+\epsilon \bar{f}r_0\sin (\theta +\bar{\delta}-\delta
_{00})-\epsilon \bar{r}_0\sin (\theta +\delta _{-+}-\bar{\delta}%
_{00})+\epsilon ^2\bar{f}\sin (\bar{\delta}-\delta _{-+})\right]  \label{49a}
\\
{\cal R}_{-}^0-{\cal R}_{+}^0 &=&\mp 4\cos \gamma \left[ r_0\bar{r}_0\cos
(\delta _{00}-\bar{\delta}_{00})+\epsilon \bar{f}r_0\cos (\theta +\bar{\delta%
}-\delta _{00})+\epsilon \bar{r}_0\cos (\theta +\delta _{-+}-\bar{\delta}%
_{00})\right] \simeq \mp 4\cos \gamma \left[ r_0\bar{r}_0\cos (\delta _{00}-%
\bar{\delta}_{00})\right]  \label{49b}
\end{eqnarray}
where 
\end{mathletters}
\begin{eqnarray}
f &=&\frac{a_{-0}}{a_{-+}}=\left( 1+C/T\right) \simeq 1.22  \nonumber \\
\bar{r} &=&\frac{\bar{a}_{-0}}{a_{-+}}=\left( \frac{\bar{C}+\bar{A}}{\bar{T}}%
\right) \frac{\bar{T}}T=\sqrt{\rho ^2+\eta ^2}\left( 0.72\right) \times 0.30
\nonumber \\
r_0 &=&\frac{a_{00}}{a_{-+}}=\frac CT=\left( a_2/a_1\right)  \nonumber \\
\bar{r}_0 &=&\frac{\bar{a}_{00}}{a_{-+}}=\frac{\bar{C}}{\bar{T}}\left( \frac{%
\bar{T}}T\right) =\sqrt{\rho ^2+\eta ^2}\left( a_2/a_1\right) \left(
0.72\right)  \nonumber \\
\bar{f} &=&\frac{\bar{a}}{a_{-+}}=\frac{\bar{T}}T=\sqrt{\rho ^2+\eta ^2}%
\left( 0.72\right)  \label{50}
\end{eqnarray}
As is clear from Eqs. (\ref{48a}) and (\ref{49a}), the FSI corrections tend
to cancel each other; in ${\cal A}_{\mp }^0$ the cancellation is almost
complete and one gets ${\cal A}_{\mp }^0\simeq 0$, where as for ${\cal A}%
_{\mp }$ one gets the value $\left( 1\times 10^{-3}\right) \sin \gamma $.
>From Eqs. (\ref{48b}) and (\ref{49b}), using Eq. (\ref{50}) and phase
shifts from TableIV (cf first column), we get 
\begin{eqnarray}
{\cal R}_{-}-{\cal R}_{+} &\simeq &\left( 0.42\right) \cos \gamma  \nonumber
\\
{\cal R}_{-}^0-{\cal R}_{+}^0 &\simeq &\left( 0.23\right) \cos \gamma
\label{51}
\end{eqnarray}
Let us now discuss the direct CP-violation for $B_s$-decay. Defining $B_{\mp
}^0=\frac 1{\sqrt{2}}\left( B_s^0\mp \bar{B}_s^0\right) $, we get using Eqs.
(\ref{28d}) and (\ref{29e}) 
\begin{mathletters}
\label{11}
\begin{eqnarray}
\frac{2\left| A\left( B_{\mp }\rightarrow K^{+}D_s^{-}\right) \right|
^2-b_s^2-\bar{b}_s^2}{2b_s\bar{b}_s} &=&\mp \left[ \cos \left( \gamma
-\delta _s+\bar{\delta}_s\right) \right]  \label{52a} \\
\frac{2\left| A\left( B_{\mp }\rightarrow K^{-}D_s^{+}\right) \right|
^2-b_s^2-\bar{b}_s^2}{2b_s\bar{b}_s} &=&\mp \left[ \cos \left( \gamma
+\delta _s-\bar{\delta}_s\right) \right]  \label{52b}
\end{eqnarray}
Since $\delta _s=\bar{\delta}_s$, it implies

\end{mathletters}
\begin{equation}
\Gamma \left( B_{\mp }\rightarrow K^{+}D_{s}^{-}\right) =\Gamma \left(
B_{\mp }\rightarrow K^{-}D_{s}^{+}\right)  \label{53}
\end{equation}
Our result $\delta _{s}=\bar{\delta}_{s}$, has important implication for
determining the phase $\gamma $ discussed in reference \cite{f15}.

Finally we discuss the time dependent analysis of $B$ decays to get
information about weak phase $\gamma $. Following the well known procedure 
\cite{f16}, \cite{f17} the time dependent decay rate for $B^0(t)$ and $\bar{B%
}^0(t)$ are given by 
\begin{eqnarray}
{\cal A}(t) &\equiv &\frac{[\Gamma _f(t)+\Gamma _{\bar{f}}(t)]-[\bar{\Gamma}%
_f(t)+\bar{\Gamma}_{\bar{f}}(t)]}{\Gamma _f(t)+\bar{\Gamma}_f(t)}  \nonumber
\label{57} \\
&\simeq &-2\frac{\sin (\Delta m_Bt)\sin (2\beta +\gamma )\times
[\left\langle f\left| H\right| B^0\right\rangle ^{*}\left\langle f\left|
H\right| \bar{B}^0\right\rangle +\left\langle f\left| H\right| \bar{B}%
^0\right\rangle ^{*}\left\langle f\left| H\right| B^0\right\rangle ]}{\left|
\left\langle f\left| H\right| B^0\right\rangle \right| ^2+\left|
\left\langle f\left| H\right| \bar{B}^0\right\rangle \right| ^2}  \label{54}
\end{eqnarray}
and 
\begin{eqnarray}
{\cal F}(t) &=&\frac{[\Gamma _f(t)+\Gamma _{\bar{f}}(t)]-[\bar{\Gamma}%
_f(t)+\Gamma _{\bar{f}}(t)]}{\Gamma _f(t)+\bar{\Gamma}_f(t)}  \nonumber \\
&=&\frac{-2}{\left| \left\langle f\left| H\right| B^0\right\rangle \right|
^2+\left| \left\langle f\left| H\right| \bar{B}^0\right\rangle \right| ^2}%
\left[ \cos (\Delta m_Bt)[\left| \left\langle f\left| H\right| \bar{B}%
^0\right\rangle \right| ^2-\left| \left\langle f\left| H\right|
B^0\right\rangle \right| ^2]\right.  \nonumber  \label{55} \\
&&\left. -i\sin (\Delta m_Bt)\cos (2\beta +\gamma )[\left\langle f\left|
H\right| B^0\right\rangle ^{*}\left\langle f\left| H\right| \bar{B}%
^0\right\rangle -\left\langle f\left| H\right| \bar{B}^0\right\rangle
^{*}\left\langle f\left| H\right| B^0\right\rangle ]\right]  \label{55}
\end{eqnarray}
where 
\begin{equation}
\Gamma _f,_{\bar{f}}(t)\equiv \Gamma \left( B^0(t)\rightarrow f,\bar{f}%
\right) ,\bar{\Gamma}_f,_{\bar{f}}(t)\equiv \Gamma \left( \bar{B}%
^0(t)\rightarrow f,\bar{f}\right)  \label{56}
\end{equation}
Taking $f\equiv K_sD^0$ and $\bar{f}\equiv K_s\bar{D}^0$ and using Eqs. (\ref
{29}) and (\ref{30}) we get 
\begin{equation}
{\cal A}(t)=-4\frac{\Gamma \left( \bar{B}^0\rightarrow K^{-}D^{+}\right)
(a_2/a_1)(C/T)\sqrt{\rho ^2+\eta ^2}}{\Gamma \left( \bar{B}^0\rightarrow 
\bar{K}^0D^0\right) +\Gamma \left( \bar{B}^0\rightarrow \bar{K}^0\bar{D}%
^0\right) }\left[ \sin (\Delta m_Bt)\sin (2\beta +\gamma )\times Y\right]
\label{57}
\end{equation}
\begin{eqnarray}
&&{\cal F}(t)+2\cos (\Delta m_Bt)\frac{\Gamma \left( \bar{B}^0\rightarrow 
\bar{K}^0D^0\right) -\Gamma \left( \bar{B}^0\rightarrow \bar{K}^0\bar{D}%
^0\right) }{\Gamma \left( \bar{B}^0\rightarrow \bar{K}^0D^0\right) +\Gamma
\left( \bar{B}^0\rightarrow \bar{K}^0\bar{D}^0\right) }  \nonumber
\label{58} \\
&=&-4\left[ \frac{\Gamma \left( \bar{B}^0\rightarrow K^{-}D^{+}\right)
(a_2/a_1)(C/T)\sqrt{\rho ^2+\eta ^2}}{\Gamma \left( \bar{B}^0\rightarrow 
\bar{K}^0D^0\right) +\Gamma \left( \bar{B}^0\rightarrow \bar{K}^0\bar{D}%
^0\right) }\sin (\Delta m_Bt)\cos (2\beta +\gamma )\times Z\right]
\label{58}
\end{eqnarray}
where 
\begin{equation}
Y=\left[ \cos (\delta _{00}-\bar{\delta}_{00})+\epsilon (a_1/a_2)\cos
(\theta +\bar{\delta}-\delta _{00})+\epsilon (a_1/a_2)\epsilon \cos (\theta
+\delta _{-+}-\bar{\delta}_{00})+\epsilon ^2(a_1/a_2)^2\cos (\bar{\delta}%
-\delta _{-+})\right]  \label{59}
\end{equation}
\begin{equation}
Z=\sin (\delta _{00}-\bar{\delta}_{00})+\epsilon \left( \frac{a_1}{a_2}%
\right) \sin (\theta +\bar{\delta}-\delta _{00})+\epsilon \left( \frac{a_1}{%
a_2}\right) \sin (\theta +\delta _{-+}-\bar{\delta}_{00})+\epsilon ^2\left( 
\frac{a_1}{a_2}\right) ^2\sin (\bar{\delta}-\delta _{-+})  \label{60}
\end{equation}
\mbox{$>$}%
>From Eqs. (\ref{57}) and (\ref{58}) it follows that the experimental
measurements of ${\cal A}(t)$ and ${\cal F}(t)$ would give $\sin (2\beta
+\gamma )Y$ and $\cos (2\beta +\gamma )Z$. Note that Eqs. (\ref{57}) and (%
\ref{58}) directly gives $\tan (2\beta +\gamma )Y/Z$. However, the relations 
$\bar{\delta}=\delta _{-+},$ $\delta _{00}=\bar{\delta}_{00}$ do not depend
on the detail of the model. In this case 
\begin{equation}
Z=2\epsilon \left( \frac{a_1}{a_2}\right) \sin \theta \approx 0.69
\label{61}
\end{equation}
\begin{equation}
Y=1+2\epsilon \left( \frac{a_1}{a_2}\right) \cos \theta +\epsilon
^2(a_1/a_2)^2\approx 1.34  \label{62}
\end{equation}
where, we have used $\epsilon =0.08$, $\theta =73^0$ and $a_2/a_1\approx
0.22 $ to give an order of magnitude for $Y$ and $Z$. It is clear that the
dominant contribution to $Z$ comes from the rescattering, where for $Y$, the
rescattering corrections are in the range of $33\%$.

For time dependent decays of $B_s^0,$ one can get 
\begin{equation}
{\cal A}_s(t)\equiv \frac{\Gamma _{f_s}(t)-\bar{\Gamma}_{\bar{f}_s}(t)}{%
\Gamma _{f_s}(t)+\bar{\Gamma}_{f_s}(t)}=\frac{b_s\bar{b}_s}{b_s^2+\bar{b}_s^2%
}\sin (\Delta m_{B_s}t)[S+\bar{S}]  \label{63}
\end{equation}
\begin{eqnarray}
{\cal F}_s(t) &\equiv &\frac{[\Gamma _{f_s}(t)+\bar{\Gamma}_{\bar{f}%
_s}(t)]-[\Gamma _{\bar{f}_s}(t)+\bar{\Gamma}_{f_s}(t)]}{\Gamma _{f_s}(t)+%
\bar{\Gamma}_{f_s}(t)}  \nonumber \\
&=&2\left[ \frac{(b_s^2-\bar{b}_s^2)\cos (\Delta m_{B_s}t)+b_s\bar{b}_s(S-%
\bar{S})\sin (\Delta m_{B_s}t)}{b_s^2+\bar{b}_s^2}\right]  \label{64}
\end{eqnarray}
where $f_s=K^{+}D_s^{-}$, $\bar{f}_s=K^{-}D_s^{+}$ and 
\begin{eqnarray}
S &=&\sin (2\phi _{Ms}+\gamma +\delta _s-\bar{\delta}_s)  \nonumber \\
\bar{S} &=&\sin (2\phi _{Ms}+\gamma -\delta _s+\bar{\delta}_s)  \label{65}
\end{eqnarray}
Since $b_s^2$ and $\bar{b}_s^2$ are given by the decay widths $\Gamma \left( 
\bar{B}_s^0\rightarrow K^{-}D_s^{+}\right) $ and $\Gamma \left( \bar{B}%
_s^0\rightarrow K^{+}D_s^{-}\right) $ respectively, it is clear from Eqs. (%
\ref{63}) and (\ref{64}) that it is possible to determine $S$ and $\bar{S}$
from the experimental value of ${\cal A}_s(t)$ and ${\cal F}_s(t)$. However
for $B_s^0$, $\phi _{Ms}\simeq 0$. But $\delta _s=\bar{\delta}_s$ follows
from general arguments (cf. Eq. (\ref{39})), it is therefore reasonable to
use $\delta _s=\bar{\delta}_s$. In this case $S=\bar{S}$ ,it is then
possible to determine $\sin \gamma $, using Eq. (\ref{63}).

Finally we give an estimate of the CP asymmetry parameter ${\cal A}(t)$ and $%
{\cal A}_s(t)$, using $\sqrt{\rho ^2+\eta ^2}=0.36$. Then we find from Eqs. (%
\ref{59}), (\ref{61}) and (\ref{63}) 
\begin{eqnarray*}
{\cal A}(t) &\approx &-4\frac{\sqrt{\rho ^2+\eta ^2}}{1+\rho ^2+\eta ^2}%
\left[ \sin (\Delta m_Bt)\sin (2\beta +\gamma )\times 1.34\right] \\
&\approx &-1.94\sin (\Delta m_Bt)\sin (2\beta +\gamma )
\end{eqnarray*}
\begin{eqnarray*}
{\cal A}_s(t) &\approx &2\frac{\sqrt{\rho ^2+\eta ^2}\bar{T}/T}{1+\left(
\rho ^2+\eta ^2\right) \bar{T}/T}\sin (\Delta m_{B_s}t)\sin \gamma \\
&\approx &0.49\sin (\Delta m_{B_s}t)\sin \gamma
\end{eqnarray*}

To conclude: The rescattering corrections are of the order of $15-20\%$;
except for $\bar{B}_s^0\rightarrow \pi ^{-}D^{+}$ and $\bar{B}%
_s^0\rightarrow \pi ^{+}D^{-}$ where they are greater than the decay
amplitude given by $W$-exchange graph. The direct $CP$-asymmetry parameter
is of the order of $10^{-3}\sin \gamma $ for $B^{-}\rightarrow K^{-}D_{\mp
}^0$ and $B^{+}\rightarrow K^{+}D_{\mp }^0$ decays as the rescattering
correction tend to cancel each other for these decays. But for the time
dependent $CP$-asymmetry we get the value $-1.94\sin (2\beta +\gamma )$ for $%
B^0\rightarrow K_sD^0$, $K_s\bar{D}^0$ decays. For $B_s^0\rightarrow K^{\pm
}D_s^{\mp }$ decays our analysis gives the strong phase shifts $\delta _s=%
\bar{\delta}_s$ and we get time dependent $CP$-asymmetry of the order $%
\left( 0\cdot 49\right) \sin \gamma $ which may be used to extract the weak
phase $\gamma $ in future experiments. Finally the formalism developed for
final state interactions in this paper is also applicable for the $\Delta
S=0,$ $\Delta C=\pm 1$ $B$--decays

\[
\begin{array}{c}
\text{{\bf Table I}: Amplitudes for }\Delta C=\pm 1\text{ and }\Delta S=-1%
\text{ decay modes of }\bar{B}. \\ 
\Delta C=+1,\Delta S=-1\text{ decays} \\ 
\begin{tabular}{|c|c|c|}
\hline
Mode & Amplitude & A$_{\text{topoloy}}$ \\ \hline
$B^{-}\rightarrow K^{-}D^{0}$ & $A_{-0}=a_{-0}e^{i\delta _{-0}}=2A_{1}$ & $%
Te^{i\delta _{T}}+Ce^{i\delta _{C}}$ \\ \hline
$\bar{B}^{0}\rightarrow K^{-}D^{+}$ & $A_{-+}=a_{-+}e^{i\delta
_{-+}}=A_{1}+A_{0}$ & $Te^{i\delta _{T}}$ \\ \hline
$\bar{B}^{0}\rightarrow \bar{K}^{0}D^{0}$ & $A_{00}=a_{00}e^{i\delta
_{00}}=A_{1}-A_{0}$ & $Ce^{i\delta _{C}}$ \\ \hline
$\bar{B}_{s}^{0}\rightarrow K^{-}D_{s}^{+}$ & $B_{-s^{+}}=b_{s}e^{i\delta
_{s}}$ & $Te^{i\delta _{T}}+Ee^{i\delta _{E}}$ \\ \hline
$\bar{B}_{s}^{0}\rightarrow \pi ^{-}D^{+}$ & $B_{-+}=b_{1/2}e^{i\delta
_{1/2}}=B_{1/2}$ & $Ee^{i\delta _{E}}$ \\ \hline
$\bar{B}_{s}^{0}\rightarrow \pi ^{0}D^{0}$ & $B_{00}=\frac{1}{\sqrt{2}}%
b_{1/2}e^{i\delta _{1/2}}=\frac{1}{\sqrt{2}}B_{1/2}$ & $\frac{1}{\sqrt{2}}%
Ee^{i\delta _{E}}$ \\ \hline
\end{tabular}
\\ 
\Delta C=-1,\Delta S=-1\text{ decays} \\ 
\begin{tabular}{|c|c|c|}
\hline
Mode & Amplitude & A$_{\text{topoloy}}$ \\ \hline
$\bar{B}^{0}\rightarrow \bar{K}^{0}D^{0}$ & $\bar{A}_{00}=\bar{a}%
_{00}e^{i\left( \bar{\delta}_{00}+\gamma \right) }=2\bar{A}_{1}$ & $\bar{C}%
e^{i\left( \bar{\delta}_{C}+\gamma \right) }$ \\ \hline
$B^{-}\rightarrow K^{-}\bar{D}^{0}$ & $\bar{A}_{-0}=\bar{a}_{-0}e^{i\left( 
\bar{\delta}_{-0}+\gamma \right) }=\bar{A}_{1}+\bar{A}_{0}$ & $\left( \bar{C}%
e^{i\bar{\delta}_{C}}+\bar{A}e^{i\bar{\delta}_{A}}\right) e^{i\gamma }$ \\ 
\hline
$\bar{B}^{0}\rightarrow \bar{K}^{0}D^{-}$ & $\bar{A}_{0-}=\bar{a}%
_{0-}e^{i\left( \bar{\delta}_{0-}+\gamma \right) }=-\bar{A}_{1}+\bar{A}_{0}$
& $\bar{A}e^{i\left( \bar{\delta}_{A}+\gamma \right) }$ \\ \hline
$\bar{B}^{0}\rightarrow \pi ^{+}D_{s}^{-}$ & $\bar{A}_{+s^{-}}=\bar{a}%
e^{i\left( \bar{\delta}+\gamma \right) }$ & $\bar{T}e^{i\left( \bar{\delta}%
_{T}+\gamma \right) }$ \\ \hline
$B^{-}\rightarrow \pi ^{0}D_{s}^{-}$ & $\bar{A}_{0s^{-}}=\frac{1}{\sqrt{2}}%
\bar{A}_{+s^{-}}$ & $\frac{1}{\sqrt{2}}\bar{T}e^{i\left( \bar{\delta}%
_{T}+\gamma \right) }$ \\ \hline
$B^{-}\rightarrow \eta _{8}D_{s}^{-}$ & $\bar{A}_{8s^{-}}=\bar{a}%
_{8}e^{i\left( \bar{\delta}_{8}+\gamma \right) }$ & $\left( \frac{1}{\sqrt{6}%
}Te^{i\bar{\delta}_{T}}-\sqrt{\frac{2}{3}}\bar{A}e^{i\bar{\delta}%
_{A}}\right) e^{i\gamma }$ \\ \hline
$\bar{B}_{s}^{0}\rightarrow K^{+}D_{s}^{-}$ & $\bar{B}_{+s^{-}}=\bar{b}%
_{s}e^{i\left( \bar{\delta}_{s}+\gamma \right) }$ & $\left( \bar{T}e^{i\bar{%
\delta}_{T}}+\bar{E}e^{i\bar{\delta}_{E}}\right) e^{i\gamma }$ \\ \hline
$\bar{B}_{s}^{0}\rightarrow \pi ^{+}D^{-}$ & $\bar{B}_{+-}=\bar{b}%
_{1/2}e^{i\left( \bar{\delta}_{1/2}+\gamma \right) }=\bar{B}_{1/2}$ & $\bar{E%
}e^{i\left( \bar{\delta}_{E}+\gamma \right) }$ \\ \hline
$\bar{B}_{s}^{0}\rightarrow \pi ^{0}\bar{D}^{0}$ & $\bar{B}_{00}=\frac{1}{%
\sqrt{2}}\bar{b}_{1/2}e^{i\left( \bar{\delta}_{1/2}+\gamma \right) }=\frac{1%
}{\sqrt{2}}\bar{B}_{1/2}$ & $\frac{1}{\sqrt{2}}\bar{E}e^{i\left( \bar{\delta}%
_{E}+\gamma \right) }$ \\ \hline
\end{tabular}
\end{array}
\]

\begin{eqnarray*}
\text{{\bf Table II}} &:&\text{{\bf \ }Scattering amplitudes for various
scattering processes as given by SU(3) [Eq. (\ref{13})]. } \\
&&\text{The last column gives the amplitudes in terms of Regge exchanges} \\
&&\left. F_{P}=iC_{P}(s/s_{0})e^{bt},\text{ \quad }F_{\rho }=\gamma _{0}^{2}%
\frac{1}{\sin \pi \alpha (t)}(s/s_{0})^{\alpha (t)},\quad C_{P}\approx 5,%
\text{ \quad }\gamma _{0}^{2}\approx 72\right. \\
&& 
\begin{tabular}{|c|c|c|}
\hline
Scattering Prpocesses & Scattering amplitude & $
\begin{array}{c}
\text{Scattering amplitude in terms of Regge amplitudes} \\ 
F_{P}\text{ and }F_{\rho }\text{ Eqs. (\ref{20}) and (\ref{21}).}
\end{array}
$ \\ \hline
$K^{-}D^{+}\rightarrow K^{-}D^{+}$ & $-\frac{2}{3}F_{2}+F_{3}$ & $F_{P}:\bar{%
K}^{0}D^{0}\rightarrow \bar{K}^{0}D^{0}$ \\ \hline
$K^{-}D^{0}\rightarrow K^{-}D^{0}$ & $F_{1}+\frac{1}{3}F_{2}+F_{3}$ & $%
F_{P}+F_{\rho }$ \\ \hline
$K^{-}D^{+}\rightleftharpoons \bar{K}^{0}D^{0}$ & $F_{1}+F_{2}$ & $F_{\rho }$
\\ \hline
$K^{-}D_{s}^{+}\rightarrow K^{-}D_{s}^{+}$ & $-F_{1}+\frac{1}{3}F_{2}+F_{3}$
& $F_{P}+e^{-i\pi \alpha (t)}F_{\rho }:\pi ^{-}D^{+}\rightarrow \pi
^{-}D^{+} $ \\ \hline
$K^{-}D_{s}^{+}\rightarrow \pi ^{-}D^{+}$ & $-F_{1}+F_{2}$ & $e^{-i\pi
\alpha (t)}F_{\rho }:\sqrt{2}\left( K^{-}D_{s}^{+}\rightarrow \pi
^{0}D^{0}\right) $ \\ \hline
$K^{-}\bar{D}^{0}\rightarrow K^{-}\bar{D}^{0}$ & $\bar{F}_{1}+\frac{1}{3}%
\bar{F}_{2}+\bar{F}_{3}$ & $F_{P}+e^{-i\pi \alpha (t)}F_{\rho }:\bar{K}%
^{0}D^{-}\rightarrow \bar{K}^{0}D^{-}$ \\ \hline
$\bar{K}^{0}D^{-}\rightleftharpoons \bar{K}^{0}D^{-}$ & $\bar{F}_{1}+\bar{F}%
_{2}$ & $e^{-i\pi \alpha (t)}F_{\rho }$ \\ \hline
$\bar{K}^{0}\bar{D}^{0}\rightarrow \bar{K}^{0}\bar{D}^{0}$ & $-\frac{2}{3}%
\bar{F}_{2}+\bar{F}_{3}$ & $F_{P}$ \\ \hline
$\pi ^{+}D_{s}^{-}\rightarrow \pi ^{+}D_{s}^{-}$ & $-\frac{2}{3}\bar{F}_{2}+%
\bar{F}_{3}$ & $F_{P}:\pi ^{0}D_{s}^{-}\rightarrow \pi ^{0}D_{s}^{-}$ \\ 
\hline
$\eta _{8}D_{s}^{-}\rightarrow \eta _{8}D_{s}^{-}$ & $\frac{2}{3}\bar{F}_{2}+%
\bar{F}_{3}$ & $F_{P}+\left( 1+e^{-i\pi \alpha (t)}\right) F_{\rho }$ \\ 
\hline
$\pi ^{+}D_{s}^{-}\rightarrow \bar{K}^{0}\bar{D}^{0}$ & $-\bar{F}_{1}+\bar{F}%
_{2}$ & $F_{\rho }$ \\ \hline
$\pi ^{0}D_{s}^{-}\rightarrow K^{-}\bar{D}^{0}$ & $\frac{1}{\sqrt{2}}\left( -%
\bar{F}_{1}+\bar{F}_{2}\right) $ & $\frac{1}{\sqrt{2}}F_{\rho }:-\left( \pi
^{0}D_{s}^{-}\rightarrow \bar{K}^{0}D^{-}\right) $ \\ \hline
$\eta _{8}D_{s}^{-}\rightarrow \bar{K}^{0}D^{-}$ & $-\frac{1}{\sqrt{6}}%
\left( 3\bar{F}_{1}+\bar{F}_{2}\right) $ & $\frac{1}{\sqrt{6}}(1-2e^{-i\pi
\alpha (t)})F_{\rho }:\eta _{8}D_{s}^{-}\rightarrow K^{-}\bar{D}^{0}$ \\ 
\hline
$K^{+}D_{s}^{-}\rightarrow K^{+}D_{s}^{-}$ & $\bar{F}_{1}+\frac{1}{3}\bar{F}%
_{2}+\bar{F}_{3}$ & $F_{P}+e^{-i\pi \alpha (t)}F_{\rho }:\pi
^{+}D^{-}\rightarrow \pi ^{+}D^{-}$ \\ \hline
$K^{+}D_{s}^{-}\rightarrow \pi ^{+}D^{-}$ & $\bar{F}_{1}+\bar{F}_{2}$ & $%
e^{-i\pi \alpha (t)}F_{\rho }:\sqrt{2}\left( K^{+}D_{s}^{-}\rightarrow \pi
^{0}\bar{D}^{0}\right) $ \\ \hline
\end{tabular}
\end{eqnarray*}

\newpage

\begin{eqnarray*}
\text{{\bf Table III}} &:&\text{ Partial wave }l=0\text{ scattering
amplitude }f\text{ for elastic scattering; }S=1+2if=\eta e^{2i\Delta } \\
&& 
\begin{tabular}{|c|c|c|c|c|c|}
\hline
Scattering process & $f$ & $S$ & $\eta $ & $\Delta $ & $\frac{1-\eta }{%
1+\eta }$ \\ \hline
$K^{-}D^0\rightarrow K^{-}D^0$ & $0.08+0.12i$ & $0.76+0.16i$ & $0.78$ & $6^0$
& $0.12$ \\ \hline
$K^{-}D^{+}\rightarrow K^{-}D^{+}$ & $0.12i$ & $0.76$ & $0.76$ & $0$ & $0.14$
\\ \hline
$\bar{K}^0D^0\rightarrow \bar{K}^0D^0$ & $0.12i$ & $0.76$ & $0.76$ & $0$ & $%
0.14$ \\ \hline
$K^{-}D_s^{+}\rightarrow K^{-}D_s^{+}$ & $0.04+0.16i$ & $0.68+0.08i$ & $0.68$
& $3.5^0$ & $0.19$ \\ \hline
$\pi ^{-}D^{+}\rightarrow \pi ^{-}D^{+}$ & $0.04+0.16i$ & $0.68+0.08i$ & $%
0.68$ & $3.5^0$ & $0.19$ \\ \hline
$\bar{K}^0\bar{D}^0\rightarrow \bar{K}^0\bar{D}^0$ & $0.12i$ & $0.76$ & $%
0.76 $ & $0$ & $0.14$ \\ \hline
$\pi ^{+}D_s^{-}\rightarrow \pi ^{+}D_s^{-}$ & $0.12i$ & $0.76$ & $0.76$ & $%
0 $ & $0.14$ \\ \hline
$\pi ^0D_s^{-}\rightarrow \pi ^0D_s^{-}$ & $0.12i$ & $0.76$ & $0.76$ & $0$ & 
$0.14$ \\ \hline
$K^{-}\bar{D}^0\rightarrow K^{-}\bar{D}^0$ & $0.04+0.16i$ & $0.68+0.08i$ & $%
0.68$ & $3.5^0$ & $0.19$ \\ \hline
$\bar{K}^0D^{-}\rightarrow \bar{K}^0D^{-}$ & $0.04+0.16i$ & $0.68+0.08i$ & $%
0.68$ & $3.5^0$ & $0.19$ \\ \hline
$
\begin{array}{c}
K^{+}D_s^{-}\rightarrow K^{+}D_s^{-}, \\ 
\pi ^{+}D^{-}\rightarrow \pi ^{+}D^{-}
\end{array}
$ & $0.04+0.16i$ & $0.68+0.08i$ & $0.68$ & $3.5^0$ & $0.19$ \\ \hline
\end{tabular}
\end{eqnarray*}

\begin{eqnarray*}
&&\left. \text{{\bf Table IV}: Final state strong interaction phase shift }%
\delta _f\text{ for }\rho =\sqrt{\frac{1-\eta }{1+\eta }},\text{ 0.5 and 1}%
\right. \\
&& 
\begin{tabular}{cccc}
phase shift in degree $\rho $ & $\sqrt{\frac{1-\eta }{1+\eta }}$ & $0.5$ & $%
1 $ \\ 
$\delta _{-0}$ & $6$ & $13$ & $25$ \\ 
$\delta _{-+}$ & $0$ & $7$ & $20$ \\ 
$\delta _{00}$ & $0$ & $7$ & $20$ \\ 
$\delta _s$ & $3$ & $10$ & $23$ \\ 
$\delta _{1/2}$ & $3$ & $10$ & $23$ \\ 
$\bar{\delta}_{00}$ & $0$ & $7$ & $20$ \\ 
$\bar{\delta}$ & $0$ & $7$ & $20$ \\ 
$\bar{\delta}_{-0}$ & $3$ & $10$ & $23$ \\ 
$\bar{\delta}_{0-}$ & $3$ & $10$ & $23$ \\ 
$\bar{\delta}_s$ & $3$ & $10$ & $23$ \\ 
$\bar{\delta}_{1/2}$ & $3$ & $10$ & $23$%
\end{tabular}
\end{eqnarray*}


\begin{references}
\bibitem{f1}  M.~Gronau and D.~Wyler, {\em On determining a weak phase from
CP asymmetries in charged b decays}, Phys. Lett. B {\bf 265} (1991), {172}.

\bibitem{f2}  D. Atwood, I Dunietz and A. Soni, {\em Enhanced CP violation
with $B\to KD^0$ ($\bar{D}^0$) modes and extraction of the CKM angle $\gamma 
$}, \prl {\bf 78} {(1997)}, {3257} [hep-ph/{9612433}].

\bibitem{f3}  M.~Gronau, {\em Weak phase gamma from color-allowed $B\to DK$
rates}, \prd {\bf 58} (1998) {037301} [hep-ph/9802315].

\bibitem{f4}  Z.-z. Xing, {\em A determination of the weak phase $\gamma $
from color-allowed $B_u^{\pm }\to DK^{\pm }$ decays}, \prd {\bf 58} (1998) {%
093005} [hep-ph/9804434].

\bibitem{f5}  M.~Gronau and J.L. Rosner, {\em Final state interaction
effects on $\gamma $ from $B\to DK$}, Phys. Lett. B{\bf 439} (1998) {171}
[hep-ph/9807447].

\bibitem{f6}  L.~Wolfenstein, {\em Parametrization of the Kobayashi-Maskawa
matrix}, \prl {\bf 51} (1983) {1945}.

\bibitem{f7}  N.~Cabibbo, {\em Unitary symmetry and leptonic decays}, \prl
{\bf 10} (1963) {531};\newline
M.~Kobayashi and T.~Maskawa, {\em CP-violation in the renormalizable theory
of weak interaction}, Prog. Theo. Phys. {\bf 49} {(1973)} {652}.

\bibitem{f8}  J.D. Bjorken, {\em Topics in B physics}, Nucl. Phys. {\bf 11}
(Proc. Suppl.) {(1989)} {325}.

\bibitem{f9}  For a review, see H.~Quinn, {\em B physics and CP-violation},
hep-ph/0111177.

\bibitem{r1}  M.~Beneke, G.~Buchalla, M.~Neubert and C.T. Sachrajda, {\em {%
QCD} factorization for $B\to \pi \pi $ decays: strong phases and
CP-violation in the heavy quark limit}, \prl {\bf 83} {(1999)} {1914}
[hep-ph/9905312]; {\em QCD factorization for exclusive, non-leptonic B meson
decays: general arguments and the case of heavy-light final states}, Nucl.
Phys. B {\bf 591}{(2000)} {313} [hep-ph/0006124].

\bibitem{r2}  C.W. Bauer, D.~Pirjol and I.W. Stewart, {\em A proof of
factorization for $B\to D\pi $}, \prl {\bf 87}{(2001)}{\ 201806}
[hep-ph/0107002].

\bibitem{f10}  M.~Suzuki and L.~Wolfenstein, {\em Final state interaction
phase in B decays}, \prd {\bf 60}{\ (1999)} {074019} [hep-ph/9903477].

\bibitem{f11}  J.F. Donoghue et al., {\em Systematics of final state
interactions in B decays}, \prl {\bf 77} {(1996)} {2187}.

\bibitem{f12}  M.~Neubert and B.~Stech, {\em Non-leptonic weak decays of b
mesons}, {\em Adv.\ Ser.\ Direct.\ High Energy Phys.} {\bf 15} (1998) 294
[hep-ph/9705292];\newline
M.~Neubert and A.A. Petrov, {\em Comments on color suppressed hadronic B
decays}, Phys. Lett. B {\bf 519} {(2001)} {50} [hep-ph/0108103].

\bibitem{f13}  A.F. Falk, A.L. Kagan, Y.~Nir and A.A. Petrov, {\em Final
state interactions and new physics in $B\to \pi K$ decays}, \prd {\bf 57} {%
(1998)} {4290} [hep-ph/9712225].

\bibitem{f14}  I. Caprini, L. Micu and C. Bourrely, {\em Dispersion
relations and rescattering effects in B nonleptonic decays}, \prd {\bf 60} {%
(1999)} {074016} [hep-ph/9904214].

\bibitem{f15}  D.~Atwood and A.~Soni, {\em Using imprecise tags of CP
eigenstates in $B_s$ and the determination of the CKM phase $\gamma $},
Phys. Lett. B {\bf 533} {(2002)} {37} [hep-ph/0112218].

\bibitem{f16}  R.~Aleksan, I.~Dunietz, B.~Kayser and F.~Le~Diberder, {\em %
CP-violation using noncp eigenstate decays of neutral B mesons}, Nucl. Phys.
B {\bf 361} {(1991)} {141};\newline
R.~Aleksan, I.~Dunietz and B.~Kayser, {\em Determining the CP-violating
phase gamma}, Z. Phys. C {\bf 54} {(1992)} {653}.

\bibitem{f17}  B.~Kayser and D.~London, {\em Exploring CP-violation with $%
B_D^0\to DK_s$ decays}, \prd {\bf 61} {(2000)} {116013} [hep-ph/9909561].
\end{references}
\end{document}